\documentclass[prl,showpacs,amsmath,amssymb,amsfonts,lengthcheck,longbibliography, nofootinbib]{revtex4-2}
\usepackage{graphicx}  % needed for figures
\usepackage{dcolumn}   % needed for some tables
\usepackage{bm}        % for math
\usepackage{braket}
\usepackage[colorlinks=true,citecolor=blue,linkcolor=blue,urlcolor=blue]{hyperref}

\usepackage{todonotes}

\begin{document}

%\title{Thermodynamic Matrix Exponentials}
\title{Thermodynamic Matrix Exponentials and Thermodynamic Parallelism}

\author{Samuel Duffield$^*$}
\affiliation{Normal Computing Corporation, New York, New York, USA}

\author{Maxwell Aifer$^*$}
\affiliation{Normal Computing Corporation, New York, New York, USA}

\author{Gavin Crooks}
\affiliation{Normal Computing Corporation, New York, New York, USA}

\author{Thomas Ahle}
\affiliation{Normal Computing Corporation, New York, New York, USA}

\author{Patrick J. Coles}
\affiliation{Normal Computing Corporation, New York, New York, USA}

\date{\today}

\begin{abstract}
Thermodynamic computing exploits fluctuations and dissipation in physical systems to efficiently solve various mathematical problems. It was recently shown that certain linear algebra problems can be solved thermodynamically, leading to a speedup scaling with the matrix dimension. The origin of this ``thermodynamic advantage" has not yet been fully explained, and it is not clear what other problems might benefit from it. Here we provide a new thermodynamic algorithm for exponentiating a real matrix. We describe a simple electrical circuit involving coupled oscillators, which can implement our algorithm. We also show that this algorithm provides an asymptotic speedup that is linear in the dimension. Finally, we introduce the concept of thermodynamic parallelism to explain this speedup, stating that thermodynamic noise provides a resource leading to effective parallelization of computations, and we hypothesize this as a mechanism to explain thermodynamic advantage more generally.
\end{abstract}

\maketitle
\def\thefootnote{*}\footnotetext{These authors contributed equally to this work.}\def\thefootnote{\arabic{footnote}}

\textit{Introduction.---}The exponential of a matrix plays a central role in the study of linear differential equations. A real matrix $A \in \mathbb{R}^{d \times d}$ specifies a homogeneous linear differential equation $dx= -A x\, dt$, whose solution may  be written
\begin{equation}
\label{matrix-exponential}
    x(t) = e^{-A t}x(0),
\end{equation}
where $e^{-A t}$ can be defined in many equivalent ways, see Table \ref{table:eA}. Examples occur in simulating classical and quantum physical systems, processing audio and video signals \cite{oppenheim1999discrete}, analyzing economic time series data~\cite{hamilton1994time}, and more recently in machine learning (e.g., reinforcement learning~\cite{lu2023structured} and large language models~\cite{gu2021efficiently}). Linear differential equations also appear in the analysis of Markov processes, whose applications include queuing theory \cite{shortle2018fundamentals}, biology \cite{allen2010introduction}, and rating systems \cite{duffield2023state}.

Due to their important and diverse applications, there has been intense research activity on the efficient computation of matrix exponentials in the last century~\cite{moler1978nineteen,moler2003nineteen}.
The most popular numerical methods are diagonalization and Padé approximation \cite{pade1892representation, al2010new}, which have time complexity $O(d^3)$ or $O(d^\omega)$ for general matrices.\footnote{Here $\omega\approx 2.371552$ is the Fast Matrix Multiplication constant~\cite{williams2023new}.} The matrix exponential can also be found by solving the initial value problem (IVP) $dx = -A x \, dt, x(0) = \hat{e}_i$ for each standard basis vector $\hat{e}_i$. This approach also requires $O(d^3)$ operations, as each IVP can be solved in $O(d^2)$ operations, but it can be parallelized by allocating one thread for each IVP,
thus taking $O(d^2)$ time if you have $d$ processors. However, parallelizing to $d$ processors can be challenging.
A modern A100 GPU has 6912 cores~\cite{pny_a100}
which allows highly parallel algorithms~\cite{chowdhury2020brief, d2020similarity}, but adding more cores is difficult due to heat dissipation and the complexity of inter-core communication.

\begin{table}[h]
    \vspace{-.99em}
    \centering
    \caption{Some ways to define and compute \( e^A \), \cite{higham2008functions}}
    \label{table:eA}
    \renewcommand{\arraystretch}{1.3} % Increase the row height
    \setlength{\tabcolsep}{7pt} % Increase spacing between columns
    \begin{tabular}{cc}
        \hline
        \noalign{\smallskip}
        Power series 
        & 
        Limit \\
        \( I + A + \frac{A^2}{2!} + \frac{A^3}{3!} + \cdots \) 
        & 
        \( \lim_{s \to \infty} \left(I + \frac{A}{s}\right)^s \) \\
        \noalign{\medskip}
        Differential system 
        & 
        Schur form \\
        \( X'(t) = AX(t), X(0) = I \) 
        & 
        %\( Z \text{diag}(e^{J_k})Z^{-1} \) \\
        \( Q \text{diag}(e^T)Q^* \) \\
        \noalign{\medskip}
        Cauchy integral 
        & 
        Padé approximation \\
        \( \frac{1}{2\pi i} \oint_{\Gamma} e^z(zI - A)^{-1} dz \) 
        &
        %\( p_{km}(A)q_{km}(A)^{-1} \)
        %(1+x/2+(3 x^2)/28+x^3/84+x^4/1680)/(1-x/2+(3 x^2)/28-x^3/84+x^4/1680) 
        %\( p_{km}(A)q_{km}(A)^{-1} \)
        \(
        (I+\frac{A}{2}
        %+\frac{3 A^2}{28}
        +\dots)
        (I-\frac{A}{2}
        %+\frac{3 A^2}{28}
        +\dots)^{-1}
        \)
        \\
        \noalign{\smallskip}
        \hline
    \end{tabular}
    \vspace{-1em}
\end{table}

\begin{figure}
\centering
  \includegraphics[width=.49\textwidth]{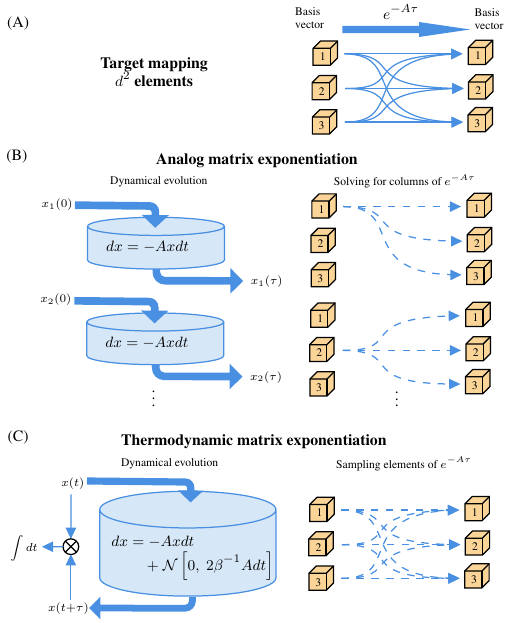}
    \caption{\textbf{Analog and thermodynamic approaches.} (A) The target matrix $e^{-A\tau}$ has $d^2$ elements that must be experimentally characterized. An individual experiment can be viewed as sampling these elements from a distribution, due to inherent randomness in the experiment (either unintentional in the analog case or intentional in the thermodynamic case). (B) A naive analog approach would sample $d$ elements at a time, while (C) The thermodynamic approach samples all $d^2$ elements at once by using noise as a resource, an advantage we call thermodynamic parallelism.}
\label{fig:AnalogComparison}
\end{figure}

An analog device, such as an electrical circuit~\cite{allen2011cmos}, can encode the (deterministic) differential equation in its voltages. However, analog devices face serious challenges due to their susceptibility to errors and the difficulty of scaling the necessary hardware to large problem sizes. For example, as illustrated in Fig.~\ref{fig:AnalogComparison}(B), implementing the matrix exponentiation method described above in an analog way would require $d$ separate analog devices, each capable of solving a $d$-dimensional linear differential equation (each device would therefore need to store the $d^2$ elements of $A$). Moreover, each device would be influenced by various types of errors, including noise caused by thermal fluctuations.
See the Supplemental Material for a full analysis of such an analog device under natural noise assumptions.

In this work, we show how to evaluate matrix exponentials on a thermodynamic computing~\cite{conte2019thermodynamic,coles2023thermodynamic,aifer2023thermodynamic,chowdhury2023full,hylton2020thermodynamic,ganesh2017thermodynamic,TherML,Boyd_2022,lipka2023thermodynamic} device with $d$ cells,
using a \emph{stochastic} differential equation (SDE). Our algorithm is not susceptible to thermal noise, and in fact relies on noise as a resource, an advantage over existing analog methods. Moreover, the stochastic dynamics of the thermodynamic device can be put into an exact correspondence with the deterministic evolution that an ensemble of $d$ copies of the device would undergo in the absence of noise. We respectively refer to these properties of our algorithm as \emph{noise resilience} and \emph{thermodynamic parallelism}, and they result in significant theoretical advantages over existing methods, both analog and digital.

Our algorithm provides a polynomial speed up over known methods, which is due to thermodynamic parallelism. While time complexity of $O(d^2)$ can also be obtained using a parallel digital or analog method (as discussed earlier), this requires physical parallelism, meaning the hardware
must be physically duplicated in order to solve the $d$ initial value problems simultaneously. Our approach allows for the computation of a matrix exponential using a single device which would otherwise be capable of solving only one IVP at a time, and the total time required scales as $O(d^2)$. As a result, a matrix exponential can be computed with similar hardware and time requirements as would be necessary to solve a single IVP using a deterministic analog device, as shown in Fig.~\ref{fig:AnalogComparison}(C). To summarize, digital and analog methods allow for a time-space tradeoff, achieving either $O(d^3)$ time with single thread or $O(d^2)$ time with $O(d)$ threads. Our thermodynamic matrix exponential algorithm runs in $O(d^2)$ time with a single device encoding $d^2$ quantities.

% \begin{figure}
% \centering
% \scalebox{0.8}{
%   \input{comparison_diag}}
%     \caption{\textbf{Analog and thermodynamic approaches.} (A) The target matrix $e^{-A\tau}$ has $d^2$ elements that must be experimentally characterized. An individual experiment can be viewed as sampling these elements from a distribution, due to inherent randomness in the experiment (either unintentional in the analog case or intentional in the thermodynamic case). (B) A naive analog approach would sample $d$ elements at a time, while (C) The thermodynamic approach samples all $d^2$ elements at once by using noise as a resource, an advantage we call thermodynamic parallelism.}
% \label{fig:AnalogComparison}
% \end{figure}

In what follows we give a basic description of the hardware necessary for this algorithm and the steps involved in its execution. We also present numerical simulation results, confirming that the time required scales no worse than $O(d^2)$. An analytical derivation of the time required for evaluating the matrix exponential can be found in the Supplemental Material.

\bigskip

\textit{Thermodynamic Matrix Exponentials.---}Let $A \in \mathbb{R}^{d \times d}$ be a real symmetric positive definite matrix (note that we will relax this assumption later in the article). Suppose we would like to find the matrix exponential $e^{-A}$. We assume that the thermodynamic device evolves a system dictated by the overdamped Langevin equation
\begin{align}
    dx = - Axdt + \mathcal{N}[0, 2\beta^{-1}B dt],
\end{align}
for scalar $\beta \in \mathbb{R}^+$ and symmetric positive definite $B$. These dynamics represent an Ornstein-Uhlenbeck (OU) process with correlation function (or correlation matrix)
\begin{equation}
    \label{correlation-matrix-def}
   C(t,s) =  \braket{[x(t) - \braket{x(t)}][x(s) - \braket{x(s)}]^\intercal}.
\end{equation}
We will assume that $\braket{x(0)}=0$, which implies that $\braket{x(t)}=0$ at all times. The correlation matrix becomes
\begin{equation*}
   C(t,s) =  \braket{x(t)x^\intercal(s)}.
\end{equation*}
At equilibrium, $x\sim \mathcal{N}[0, \Sigma]$ is normally distributed, with a correlation function given by \cite{gardiner1985handbook}
\begin{equation*}
 C(t+\tau,t)=   \braket{x(t+\tau)x(t)^\intercal} = e^{-A\tau}\Sigma.
\end{equation*}
The covariance matrix $\Sigma$ satisfies the Lyapunov equation
\begin{equation}\label{eq:lyapunov}
    A\Sigma + \Sigma A^\intercal = 2 \beta^{-1} B.
\end{equation}
Setting $B = A$ results in the solution $\Sigma = \beta^{-1}\mathbb{I}$. Therefore the SDE
\begin{equation}\label{eq:specific_OU}
    dx = -Ax dt + \mathcal{N}[0, 2 \beta^{-1} A dt]
\end{equation}
has correlation function
\begin{equation}
\label{exponential-correlation}
    C(t+\tau,t) = \beta^{-1}e^{-A\tau}.
\end{equation}
Importantly, according to Eq. \eqref{exponential-correlation} the correlation function evolves in the same way as would a collection of $d$ deterministic analog devices, which underlies the parallelism of the thermodynamic algorithm.

%\begin{align}
%dx &= - Ax dt + \mathcal{N}[0, %2 \beta^{-1} A dt]\\
%e^{-A\tau} &= \beta \langle x(t+\tau)x(t)^\intercal \rangle
%\end{align}

%Evaluating the correlation function then yields the matrix exponential.

%\begin{align}
%dx &= -Ax dt + %\mathcal{N}[0, 2 %\beta^{-1} A dt]\notag\\
% e^{-A\tau} &=\beta \langle %x(t+\tau)x(t)^\intercal \rangle \notag
%\end{align}

However, the correlation function is not directly accessible by measurement, and must be estimated via an average. Due to the ergodicity of this system~\cite{aifer2023thermodynamic, gardiner1985handbook}, we can use a time-average instead of an ensemble average, and obtain
\begin{equation}
\label{eq:exp-time-integral}
    e^{-A \tau} \approx \frac{\beta}{T} \int_{t_0}^{t_0 + T} dt \, x(t+\tau)x^\intercal(t).
\end{equation}

Thus if we can evolve the SDE in~\eqref{eq:specific_OU} (initiated at $x(t_0) = 0$) and collect samples $\{x(t_n)\}_{n=1}^N$ regularly dispersed by time $\tau$ we can extract an asymptotically unbiased approximation to $e^{-A\tau}$ with the sample autocovariance
\begin{equation}
    \frac{\beta}{N}\sum_{n=1}^N x(t_n) x(t_{n-1})^\intercal.
%    \frac{1}{N}\sum_{n=1}^N x_{n} x_{n -1}^\intercal.
\end{equation}
The parameters $\beta$ and $\tau$ are tuning parameters and the matrix exponential $e^{-A}$ can be extracted by rescaling the input $A$ by $\tau^{-1}$.

\bigskip

\textit{Arbitrary Matrices.---}We can relax the constraint that $A$ needs to be symmetric positive definite. Suppose we want $e^{-M}$ for arbitrary $M$. Then consider $A = c \mathbb{I} + M$ and $B = \frac{1}{2}(A + A^\intercal)$ (which similarly results in $\Sigma = \beta^{-1}\mathbb{I}$ solving the Lyapunov equation in~\eqref{eq:lyapunov}). The parameter $c \in \mathbb{R}^+$ must be large enough that the eigenvalues of $A$ have positive real part, which will also ensure that $B$ is positive definite and a valid diffusion matrix. The SDE becomes
\begin{equation}
    dx = -Ax dt + \mathcal{N}[0, \beta^{-1} (A + A^\intercal) dt].
\end{equation}
Applying the previous procedure will produce an approximation to
\begin{equation*}
e^{-A} = e^{-c \mathbb{I} - M} = e^{-c} e^{-M}.
\end{equation*}
Therefore a scalar rescaling by $e^c$ provides the desired $e^{-M}$. These arguments show that our thermodynamic algorithm can be used to obtain the exponential of any matrix in principle. In practice, one would have to be careful with precision if $c$ is required to be too large. We find that the time required to evaluate the matrix exponential depends on the condition number, and the condition number of $A$ differs from that of $M$. Also, if $c$ is large then there will be a large loss in precision because the elements of $e^{-A}$ will be very small, so it may be necessary to rescale $M$ as well. A task for future work is to carefully take these considerations into account in order to fully describe the complexity of the matrix exponential algorithm in the case of arbitrary matrices.

\bigskip

\textit{Thermodynamic parallelism.---}Let us now reflect on the mechanism of speedup. The advantage of the thermodynamic algorithm is that it allows for the collection of a sample of the matrix exponential in constant time (that is, time not scaling with dimension) using a device with $d$ nodes, and $d^2$ couplings between them. Using a deterministic analog algorithm, the same device could be used to solve the $d$ initial value problems in serial to obtain a single sample of the matrix exponential, resulting in an $O(d^3)$ time cost to collect the $O(d^2)$ samples.

The ability to solve $d$ initial value problems simultaneously, using a device that ostensibly only solves one, can be attributed to the probabilistic nature of the thermodynamic algorithm. Suppose that, once equilibrium has been reached, the state of the system is a random variable with an isotropic normal distribution, $x(0) \sim \mathcal{N}[0, \beta^{-1}\mathbb{I}]$. As discussed above, the correlation function is then given by $C(t+\tau, t) = \beta^{-1}e^{-A\tau}$. An analogy can be drawn to an ensemble of $d$ separate deterministic analog devices, each of which solves the differential equation $dx = - A x \,dt$. We describe the state of this system by a matrix $X$, each of whose columns is the state of a single device in the ensemble. Initially we set the state of the $i$th device to the standard basis vector $\hat{e}_i$, so $X(0) = \mathbb{I}$. It is then apparent that $X(\tau) = e^{- A \tau}$, which is (up to a constant) the same as the time-evolution of the correlation matrix of the thermodynamic device. While the state of the system $x(t)$ is a vector with only $d$ components, the correlation matrix (which is a property of the \emph{distribution}, rather than  the state) has $d^2$ elements, representing the desired matrix exponential.

Thermodynamic parallelism stems from the fact that the probability distribution of the system is more informative than the instantaneous state; this aligns with the intuition of the system exploring multiple trajectories at the same time, which is why we refer to this advantage as a form of parallelism. Without noise, the above reasoning would break down. The dynamics of the differential equation $dx = -A x dt$ are contractive and the Shannon entropy of the system tends to zero over time, meaning the correlations will eventually be too small to measure. Thus, noise can be viewed as the key resource that enables thermodynamic parallelism.

\begin{figure}
\centering
  \includegraphics[width=.35\textwidth]{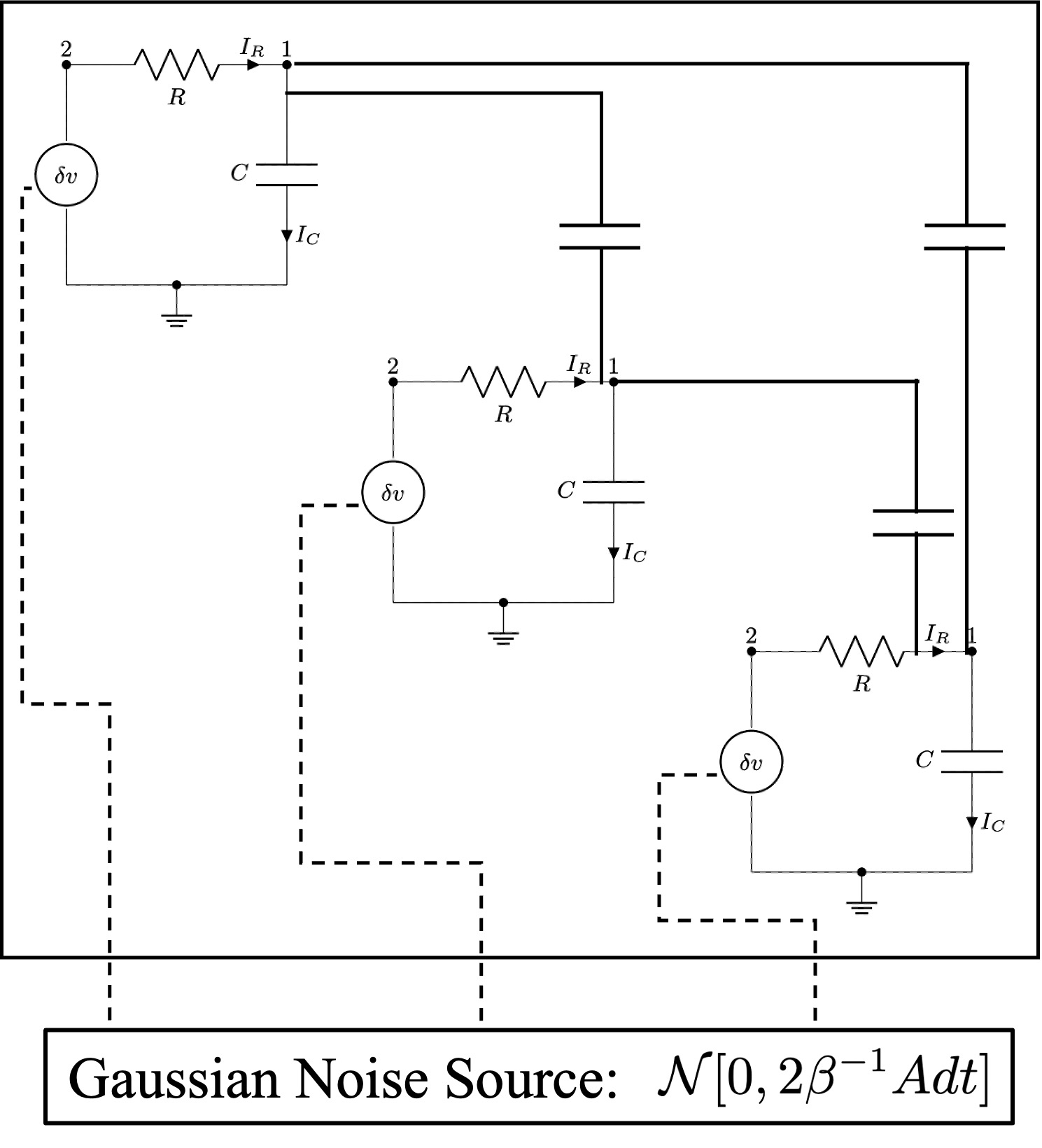}
    \caption{\textbf{Possible hardware architecture for matrix exponentiation.} The overall system is composed of two subunits. One subunit (bottom) outputs, as a vector of voltages, Gaussian noise with zero mean and covariance matrix proportional to $A$. The other subunit (top) takes this output as its noise source and directly simulates the stochastic differential equation in~\eqref{eq:specific_OU}.}
\label{fig:Circuit_Main}
\end{figure}

While we focus here on the matrix exponential algorithm, similar arguments based on thermodynamic parallelism can be used to explain the asymptotic advantage achieved by the thermodynamic matrix inversion algorithm in \cite{aifer2023thermodynamic}. In that work, the matrix inverse is found by estimating the covariance matrix of the thermodynamic device, $\Sigma = \braket{x x^\intercal}$. For a symmetric positive definite matrix $A$, the SDE $dx = - A x + \mathcal{N}[0,2\beta^{-1}\mathbb{I} dt]$ eventually reaches a stationary distribution $x\sim\mathcal{N}[0,\beta^{-1}A^{-1}]$, so $\beta \Sigma = A^{-1}$ at equilibrium \cite{aifer2023thermodynamic}. The matrix inverse could also be found by solving $d$ linear systems of equations of the form $A x = \hat{e}_i$, where $\hat{e}_i$ is the $i$th standard basis vector. An analog device could solve a linear system of this form by evolving the ODE $dx = - (A x - \hat{e}_i)\,dt$ for sufficiently long. A collection of $d$ such devices could solve the $d$ linear systems in parallel via the matrix ODE
\begin{equation}
\label{matrix-inverse-ode}
    dX = - A X \, dt+ \mathbb{I}\, dt.
\end{equation}
In the thermodynamic matrix inverse algorithm, the covariance matrix evolves under the ODE
\begin{equation}
    \label{thermo-matrix-inverse-ode}
    d\Sigma = - A \Sigma\,dt - \Sigma A \,dt+ 2 \mathbb{I}\,dt,
\end{equation}
where we set $\beta = 1$. This is a symmetrized version of Eq. \eqref{matrix-inverse-ode}, and $A$ and $\Sigma$ are symmetric, Eqs. \eqref{matrix-inverse-ode} and \eqref{thermo-matrix-inverse-ode} have the same stationary solution $X = \Sigma = A^{-1}$. This means that the thermodynamic matrix inverse algorithm allows a single device similar to one in Fig. \ref{fig:Circuit_Main} to emulate a collection of $d$ deterministic analog devices. Moreover, this suggests that thermodynamic parallelism, arising from stochastic noise, may be a broader mechanism to explain the potential advantage of thermodynamic computers.

\bigskip

\textit{Physical device construction.---}Let us discuss how one could construct a device to implement the thermodynamic matrix exponential algorithm. Essentially, we need a device that can implement the SDE in Eq.~\eqref{eq:specific_OU}. Implementing this OU process with electrical circuits has, to some extent, been discussed in Refs.~\cite{coles2023thermodynamic,aifer2023thermodynamic} where stochastic units (s-units) were proposed as the basic building block of the hardware, with each s-unit composed of an RC circuit with a stochastic voltage source. Indeed, an array of s-units, capacitively coupled to one another, is shown in the top of Fig.~\ref{fig:Circuit_Main}, and in theory this s-unit array can simulate Eq.~\eqref{eq:specific_OU}.

However, an additional subtlety about Eq.~\eqref{eq:specific_OU} is that the noise source is correlated. Hence, one needs a Gaussian noise source that outputs the vector of correlated noise values to provide the noise term in Eq.~\eqref{eq:specific_OU}. This is shown simply as a black box at the bottom of Fig.~\ref{fig:Circuit_Main}, while we give a detailed discussion on how to construct this noise source in the Supplemental Material. Namely, this noise source can consist of an (additional) array of s-units that are capactively coupled, and the voltage vector outputted by the noise source is the output of integrators that integrate the current flow through the resistors in the s-units (see Supplemental Material for details).

\begin{figure}
\centering
    \includegraphics[width=.49\textwidth]{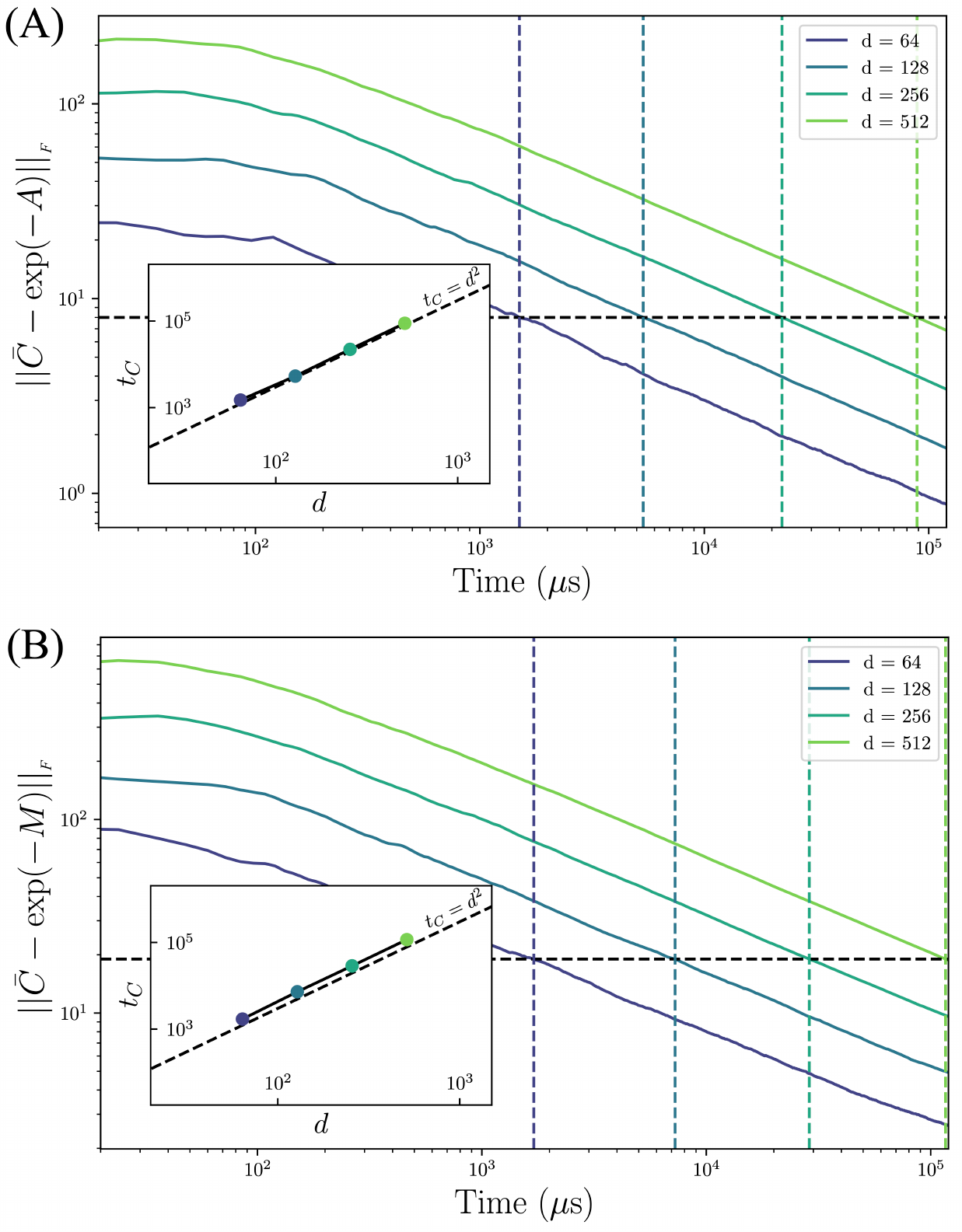}
    \caption{\textbf{Error as a function of analog integration time for varying matrix dimension.} The error is the Frobenius norm of the sampled matrix minus the true matrix exponential. (A) Matrices are positive-definite and drawn from a Wishart distribution (with $2d$ degrees of freedom). (B) Matrices are asymmetric, with elements drawn from the Haar distribution over orthogonal matrices. To make the eigenvalues have positive real part, a term $c\mathbb{I}$ is added as described in the text, with $c=1.1$. Vertical lines represent the times $t_C$ when the error falls below a threshold. Inset: Crossing time $t_C$ versus matrix dimension $d$.} 
\label{fig:results}
\end{figure}

\bigskip

\textit{Complexity.---}As outlined earlier, the thermodynamic matrix exponential algorithm has two steps:
\begin{enumerate}
    \item Wait for time $t_0$ large enough that the system is at equilibrium or initiate at a sample from the stationary distribution $x(0) \sim \mathcal{N}[0, \beta^{-1}\mathbb{I}]$ (giving $t_0{=}0$).
    \item Evaluate Eq. \eqref{eq:exp-time-integral} by integrating from time $t_0$ to time $t_0 + T$, where the integration time $T$ is determined using Eq. \eqref{eq:integration-time} below.
\end{enumerate}
We denote the time-average appearing in Eq. \eqref{eq:exp-time-integral} by $\overline{C}$, and define the RMS error
% \begin{equation*}
%      \mathcal{E} = \sqrt{\left\langle\left\| e^{-A \tau} - \beta \overline{C}\right\|_F^2\right\rangle},
% \end{equation*}
$
     \mathcal{E} = \sqrt{\left\langle\left\| e^{-A \tau} - \beta \overline{C}\right\|_F^2\right\rangle},
$
where $\|\cdot\|_F$ is the Frobenius norm. As shown in the Supplemental Material, we can achieve RMS error of $\mathcal{E}$ or lower by setting the integration time larger than the following lower bound
\begin{equation}
    \label{eq:integration-time}
    T \geq \frac{2 d(d+1) e^{2\alpha_\text{min} \tau}\kappa }{\|A\|  \mathcal{E}^2},
\end{equation}
% \begin{equation}
%     % \label{eq:integration-time}
%     T \geq \frac{2 d(d+1) e^{2\alpha_\text{min} \tau} }{\alpha_\text{min}  \mathcal{E}^2},
% \end{equation}
where $\kappa$ and $\alpha_\text{min}$ are respectively the condition number and the smallest singular value of $A$. Therefore the time complexity scaling is $O(d^2 \kappa \mathcal{E}^{-2})$ which provides a speedup for well conditioned matrices.

\bigskip
\textit{Numerics.---}Figure~\ref{fig:results} shows the numerical simulation results for our thermodynamic matrix exponentiation algorithm. The Frobenius error is plotted versus time for various dimensions, with the insets showing the time to reach a given error threshold versus dimension. For both positive-definite and more general matrices, respectively shown in Fig.~\ref{fig:results}(A) and (B), the run time does not grow faster than $O(d^2)$, as expected (as we have $\tau$ and  $\alpha_\text{min} = \frac{\kappa}{\| A\|}$ constant). This result holds for matrices drawn from a Wishart distribution and from an orthogonal Haar distribution, which provides evidence that our bounds on polynomial scaling are widely applicable.

\bigskip

\textit{Discussion.---}We have shown that matrix exponentiation can be solved in an amount of time proportional to the square of the dimension, whereas previously the best known upper estimates of asymptotic time complexity had higher-order polynomial scaling with dimension. The promise of quantum computing has often been attributed to the (still controversial) idea of quantum parallelism \cite{jozsa1991characterizing} and quantum matrix exponentiation has been proposed~\cite{lloyd2014quantum}, but the practicality and commercial impact of quantum computing remains long term in timescale. We have found that a computationally useful form of parallelism can be achieved within a classical probabilistic setting with near-term hardware, and its origin is explained unambiguously.

Matrix exponentiation is relevant to virtually all time-dependent processes that have linear feedback. These appear, of course, in physics, but also in disparate fields such as machine learning, biology, and economics. The future impact of these results, then, is limited only by the scale of thermodynamic hardware that can be built. These results also reveal a potential polynomial separation between the digital computing and thermodynamic computing paradigms, in terms of time complexity.
% As it is not yet known whether analog computers can provide asymptotic speedups over their digital counterparts in a clear way, this finding could perhaps change the landscape of computational complexity theory.

Demonstrating ``thermodynamic advantage'' over standard digital hardware can only be convincingly proven through experimentation, so a physical demonstration is a key future direction. We anticipate that, as a result of the potential advantages we have found, thermodynamic computing will become a rich and competitive ground for both theoretical and experimental work in the near future.

\bibliographystyle{unsrt}

\bibliography{references,thermo,bib}

\newpage
\pagebreak
\onecolumngrid

\begin{appendix}

\begin{center}
\Large Supplemental Material for\\ ``Thermodynamic Matrix Exponentials and Thermodynamic Parallelism''
\end{center}

\section{Overview}

In this Supplemental Material, we cover the following topics:
\begin{itemize}
    \item Thermodynamic Matrix Exponential Algorithm
    \item Convergence of the Matrix Exponential
    \item Scaling of the Relative Error
    \item Thermodynamic Noise Generation
    \item Comparison of Analog Algorithms with Thermodynamic Algorithms
    \item Thermodynamic Parallelism
\end{itemize}

\section{Thermodynamic Matrix Exponential Algorithm}

This section describes the thermodynamic algorithm for matrix inversion in more detail, and shows that the correct result is obtained. The basis of the matrix exponential algorithm is a physical system whose dynamics are governed by the Langevin equation
\begin{equation}
    \label{langevin-equation}
    dx = - Axdt + \mathcal{N}[0, \beta^{-1}(A + A^\intercal) dt],
\end{equation}
where $\beta$ is a positive real parameter and $A$ is a matrix whose eigenvalues have positive real part. A solution $x(\tau)$ can be written formally by integrating Eq. \eqref{langevin-equation}
\begin{equation}
    \label{langevin-formal-solution}
    x(\tau) = e^{-A \tau} x(0)  + \int_0^\tau dt\, e^{-A(\tau-t)} \sqrt{\beta^{-1}(A + A^\intercal)} dW.
\end{equation}
The square root above is understood as representing the left factor in the Cholesky decomposition, that is $X \equiv \sqrt{X} (\sqrt{X})^\intercal$ for any real symmetric positive definite $X$. First note that the second term in Eq. \eqref{langevin-formal-solution} has zero mean because the Brownian increment $dW$ has zero mean, so
\begin{equation}
    \label{langevin-mean}
    \braket{x(\tau)} = e^{-A \tau}\braket{x(0)}.
\end{equation}
We suppose that the initial condition is always chosen to be the zero vector $x(0) = 0$. The correlation matrix is \cite{gardiner1985handbook}
\begin{align}
   C(t,s) &=  \braket{[x(t) - \braket{x(t)}][x(s) - \braket{x(s)}]^\intercal} \\
   & = \braket{x(t) x^\intercal(s)}\\
   & = \int_0^{\text{min}(s,t)} dt' \, e^{-A(t-t')} \beta^{-1}(A + A^\intercal) e^{-A^\intercal (s - t')}.
\end{align}
For large $t$, a stationary distribution is reached \cite{gardiner1985handbook}, which is Gaussian, having mean $0$ and covariance matrix $\Sigma$, which is the unique solution to 
\begin{equation}
   \left(A\Sigma + \Sigma A^\intercal\right) = \beta^{-1}(A + A^\intercal).
\end{equation}
We see that the above is satisfied when $\Sigma = \beta^{-1}\mathbb{I}$, which is therefore the unique solution. When $t$ is large enough that the stationary distribution has been reached and $t \geq s$, the correlation matrix can be expressed as \cite{gardiner1985handbook}
\begin{equation}
    C(t,s) = e^{-A(t-s)}\Sigma  = \beta^{-1} e^{-A(t-s)}.
\end{equation}
Therefore, in order to obtain the matrix exponential $e^{-A \tau}$, we must either allow the system to come to equilibrium by waiting until time $t_0$ or initiate which a digitally sample from the equilibrium distribution $x(0) \sim \mathcal{N}[0, \beta^{-1}\mathbb{I}]$ allowing $t_0=0$. Then estimate the correlation matrix $C(t + \tau, t)$ by averaging for a length of time $T$,
\begin{equation}
    e^{-A \tau} \approx \frac{\beta}{T} \int_{t_0}^{t_0 + T} dt \, x(t + \tau)x^\intercal(t).
\end{equation}
The time necessary to compute the time average (after equilibrium has been reached) will be $\tau + T$, so the total time necessary is $t_0 + \tau + T$.

\section{Convergence of the Matrix Exponential}

In this section, we derive an equation for the integration time $T$ needed to obtain a given root-mean-square error $\mathcal{E}$. We denote the time-averaged correlation matrix used above by $\overline{C}$,
\begin{equation}
    \overline{C} =  \frac{1}{T} \int_{t_0}^{t_0 + T} dt x(t)x^\intercal(t+\tau).
\end{equation}
We compute the variance of the matrix elements of $\overline{C}$
\begin{align}
    \text{Var}(\overline{C_{ij}}) = \braket{\overline{C_{ij}}^2} - C_{ij}(t,t+\tau)^2.
\end{align}
The first term is
\begin{equation}
    \braket{\overline{C_{ij}}^2} = \frac{1}{T^2} \int_{t_0}^{t_0 + T} dt' \int_{t_0}^{t_0 + T} dt \Braket{x_i(t) x_j(t + \tau)x_i(t') x_j(t' + \tau)},
\end{equation}
and using Isserlis's theorem \cite{koopmans1995spectral}, we find
\begin{align}
   \Braket{x_i(t) x_j(t + \tau)x_i(t') x_j(t' + \tau)} &= \Braket{x_i(t) x_j(t + \tau)}\Braket{x_i(t') x_j(t' + \tau)} + \Braket{x_i(t)x_i(t')  }\Braket{x_j(t + \tau) x_j(t' + \tau)}  \\
   &+ \Braket{x_i(t) x_j(t' + \tau)}\Braket{ x_j(t + \tau) x_i(t')} \\
   & = C_{ij}(t, t+\tau)C_{ij}(t',t'+\tau) + C_{ii}(t,t')C_{jj}(t+\,t''+t) \\
   &+ C_{ij}(t',t''+ \tau) C_{ij}(t',t+\tau)
   \\
   & = C_{ij}(t, t+\tau)^2 + C_{ii}(t,t')C_{jj}(t,t') + C_{ij}(t,t'+\tau) C_{ij}(t,t'-\tau).
\end{align}
Upon plugging back in to the expression for the variance, we obtain
\begin{align}
\text{Var}(\overline{C_{ij}}) = \frac{1}{T^2} \int_{t_0}^{t_0 + T} dt' \int_{t_0}^{t_0 + T} dt C_{ii}(t,t')C_{jj}(t,t') + C_{ij}(t,t'+\tau) C_{ij}(t,t'-\tau)
\end{align}
We sum over $i$ and $j$, and derive the following upper bound 
\begin{align}
\sum_{ij}\text{Var}(\overline{C_{ij}}) &= \frac{1}{T^2} \int_{t_0}^{t_0 + T} dt' \int_{t_0}^{t_0 + T} dt\sum_{ij}C_{ii}(t,t')C_{jj}(t,t') + C_{ij}(t,t'+\tau) C_{ij}(t,t'-\tau)\\
& \leq 
\frac{1}{\tau^2} \frac{1}{T^2} \int_{t_0}^{t_0 + T} dt' \int_{t_0}^{t_0 + T} dt \sum_{ij} C_{ii}(t,t')C_{jj}(t,t') +\max\{ \|C(t,t'+\tau)\|_F^2, \| C(t,t'-\tau)\|_F^2\}\\
\end{align}
As the eigenvalues of $A$ have positive real part, all eigenvalues of $C(t,t'-\tau)$ decrease in modulus as $\tau$ increases, so the Frobenius norm decreases too. We therefore have

\begin{align}
\sum_{ij}\text{Var}(\overline{C_{ij}} )&\leq  \frac{1}{T^2} \int_{t_0}^{t_0 + T} dt' \int_{t_0}^{t_0 + T} dt \sum_{ij} C_{ii}(t,t')C_{jj}(t,t') + \|C(t,t'-\tau)\|_F^2
\end{align}
Note that $\sum_{ij}C_{ii}C_{jj} = \text{tr}(C)^2 \leq \text{tr}(\left|C\right|)^2$, which is square of the Schatten one norm. The Schatten one norm (trace norm), Schatten two norm (Frobenius norm), and spectral norm, obey the following inequalities for any matrix $X$
\begin{equation}
    \|X\| \leq \|X\|_F \leq \text{tr}(\left|X\right|)\leq \sqrt{d} \|X\|_F \leq d \|X\|.
\end{equation}
Therefore,
\begin{align}
\sum_{ij}\text{Var}(\overline{C_{ij}} )&\leq  \frac{1}{T^2} \int_{t_0}^{t_0 + T} dt' \int_{t_0}^{t_0 + T} dt d^{2}\|C(t,t')\|^2 +d \|C(t,t'-\tau)\|^2.
\end{align}
Note that for some class of problems of interest (in other words some particular set of matrices), there may be a parameter $r$, with $0 \leq r \leq 1$, such that we always have $\text{tr}(|X|)\leq d^r\|X\|$, which could give a better dependence on $d$ in the final result. Once again, we use the fact that all eigenvalues of $C(t,t'-\tau)$ are decreasing with $\tau$ to obtain the inequality
\begin{align}
\sum_{ij}\text{Var}(\overline{C_{ij}} )&\leq  \frac{1}{T^2} \int_{t_0}^{t_0 + T} dt' \int_{t_0}^{t_0 + T} dt d^{2}\|C(t,t'-\tau)\|^2 +d \|C(t,t'-\tau)\|^2\\
& = 
\frac{d(d+1)}{T^2} \int_{t_0}^{t_0 + T} dt' \int_{t_0}^{t_0 + T} dt \|C(t,t'-\tau)\|^2.
\end{align}
Next, we use the fact that in the stationary distribution for $t\geq s$, 
$C(t,s)   =  e^{-A(t-s)}\Sigma$, and that $C(t,s) = C(s,t)^\intercal$ \cite{gardiner1985handbook}. In our case, this implies that $\|C(t,t'-\tau)\|   = \beta^{-1} \|e^{-A|t'-\tau -t|}\|$, so
\begin{align}
\sum_{ij}\text{Var}(\overline{C_{ij}} )&\leq \frac{d(d+1)}{\beta^2T^2} \int_{t_0}^{t_0 + T} dt' \int_{t_0}^{t_0 + T} dt \left\|e^{-A|t'-\tau -t|}\right\|^2\\
& \leq 
\frac{d(d+1)}{\beta^2T^2} \int_{t_0}^{t_0 + T} dt' \int_{t_0}^{t_0 + T} dt e^{-2\alpha_\text{min}\left|t' - \tau - t \right|} \\
& \leq 
\frac{d(d+1)e^{2\alpha_\text{min}\tau}}{\beta^2T^2} \int_{t_0}^{t_0 + T} dt' \int_{t_0}^{t_0 + T} dt e^{-2\alpha_\text{min}\left|t'  - t \right|} \\
&  =
\frac{2d(d+1)e^{2\alpha_\text{min}\tau}}{\beta^2T^2} \int_{t_0}^{t_0 + T} dt' \int_{t_0}^{t'} dt e^{-2\alpha_\text{min}\left(t'  - t \right)} \\
&  =
\frac{2d(d+1)e^{2\alpha_\text{min}\tau}}{\beta^2T^2} \int_{t_0}^{t_0 + T} dt' \frac{1-e^{-2\alpha_\text{min}(t'-t_0)}}{2\alpha_\text{min}} \\
&  \leq
\frac{2d(d+1)e^{2\alpha_\text{min}\tau}}{\beta^2\alpha_\text{min}T}
\end{align}
Therefore the RMS Frobenius error is bounded as
\begin{align}
    \mathcal{E} &= \sqrt{\left\langle\left\| e^{-A \tau} - \beta \overline{C}\right\|_F^2\right\rangle}\\
 &= 
    \sqrt{\sum_{ij}\text{Var}(\beta\overline{C_{ij}})}\\
    & \leq \sqrt{\frac{2d(d+1)e^{2\alpha_\text{min}\tau}}{\alpha_\text{min}T}}
\end{align}
Note that this result implies that RMS error can be bounded by a constant if the integration time $T$ is increased in proportion with $d^2$. We can also express this in terms of the condition number $\kappa = \alpha_\text{max}/\alpha_\text{min}$ and the operator norm $\|A\|=\alpha_\text{max}$ to obtain
\begin{equation}
\mathcal{E} \leq \sqrt{\frac{2d(d+1)e^{2\alpha_\text{min}\tau}\kappa}{\|A\| T}},
\end{equation}
so the required integration time is (to leading order in $d$)
\begin{equation}
T \geq \frac{2 d^2 e^{2\alpha_\text{min} \tau}\kappa }{\|A\|  \mathcal{E}^2}.
\end{equation}
Note that when $\alpha_\text{min}$ is small (which is often the case in high dimensions) and $\tau \approx O(1)$, we can make the approximation $e^{2 \alpha_\text{min}\tau}\approx 1$, so
we would have integration time
\begin{equation}
T \geq \frac{2 d^2 \kappa }{\|A\|  \mathcal{E}^2}.
\end{equation}

\section{Scaling of the Relative Error}

\begin{figure}[t]
\centering
    \includegraphics[width=.99\textwidth]{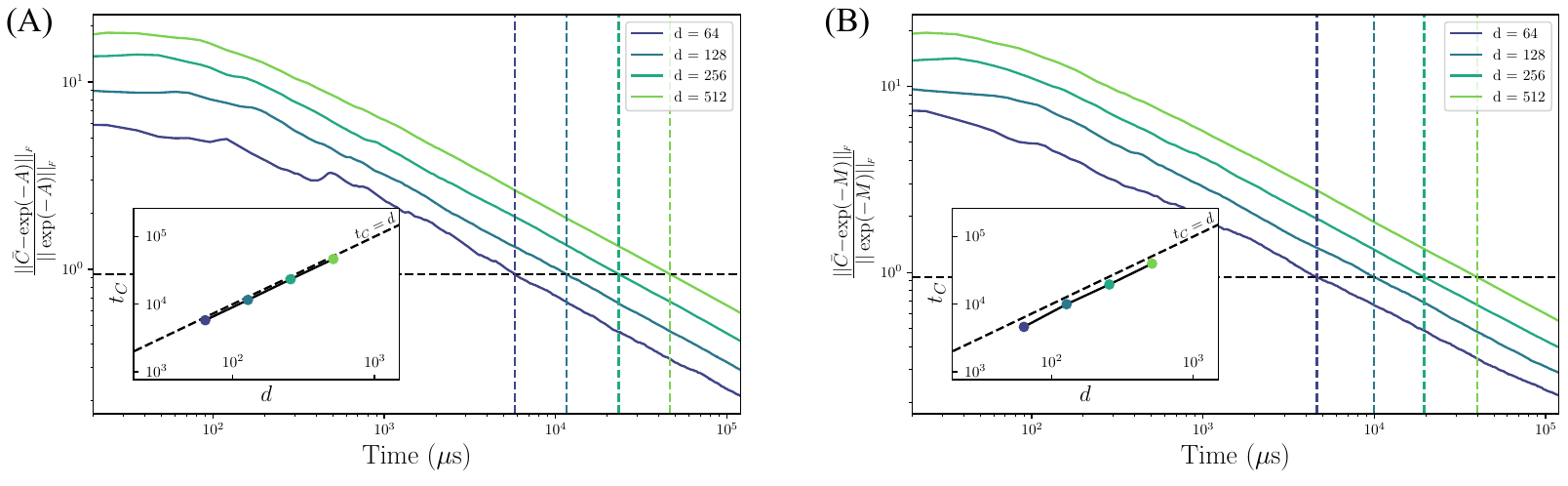}
    \caption{\textbf{Relative error as a function of analog integration time for varying matrix dimension.} The error is the relative Frobenius norm of the sampled matrix from the true matrix exponential. (A) Matrices are positive-definite and drawn from a Wishart distribution (with $2d$ degrees of freedom). (B) Matrices are asymmetric, with elements drawn from the Haar distribution over orthogonal matrices. To make the eigenvalues have positive real part, a term $c\mathbb{I}$ is added as described in the text, with $c=1.1$. Vertical lines represent the times $t_C$ when the error falls below a threshold. Inset: Crossing time $t_C$ versus matrix dimension $d$, with the dashed line showing $t_C = d$.} 
\label{fig:results_RelError}
\end{figure}

In the main text, we presented the scaling of the absolute error with dimension, for our thermodynamic matrix exponential algorithm. Here we provide the corresponding plot for relative error instead of absolute error, where relative error involves dividing the absolute error by the norm of the matrix exponential. This is shown in Fig.~\ref{fig:results_RelError}. One can see that the relative error scales more weakly with dimension as compared to the absolute error, for the class of matrices considered. For the matrices we considered, the absolute error scales as $d^2$, whereas one can see that the relative error scales approximately linearly with $d$.

\section{Thermodynamic Noise Generation}

%\begin{figure}
%\centering
%  \includegraphics[width=.40\textwidth]%{Circuit_Integrator.png}
%    \caption{\textbf{Possible hardware architecture for matrix exponentiation.} The overall system is composed of two subunits. One subunit outputs, as a vector of voltages, Gaussian noise with zero mean and covariance matrix proportional to $A$. The other subunit takes this output as its noise source and directly simulates the stochastic differential equation in~\eqref{eq:specific_OU}.}
%\label{fig:Circuit_Integrator}
%\end{figure}

\begin{figure}
\centering
  \includegraphics[width=.40\textwidth]{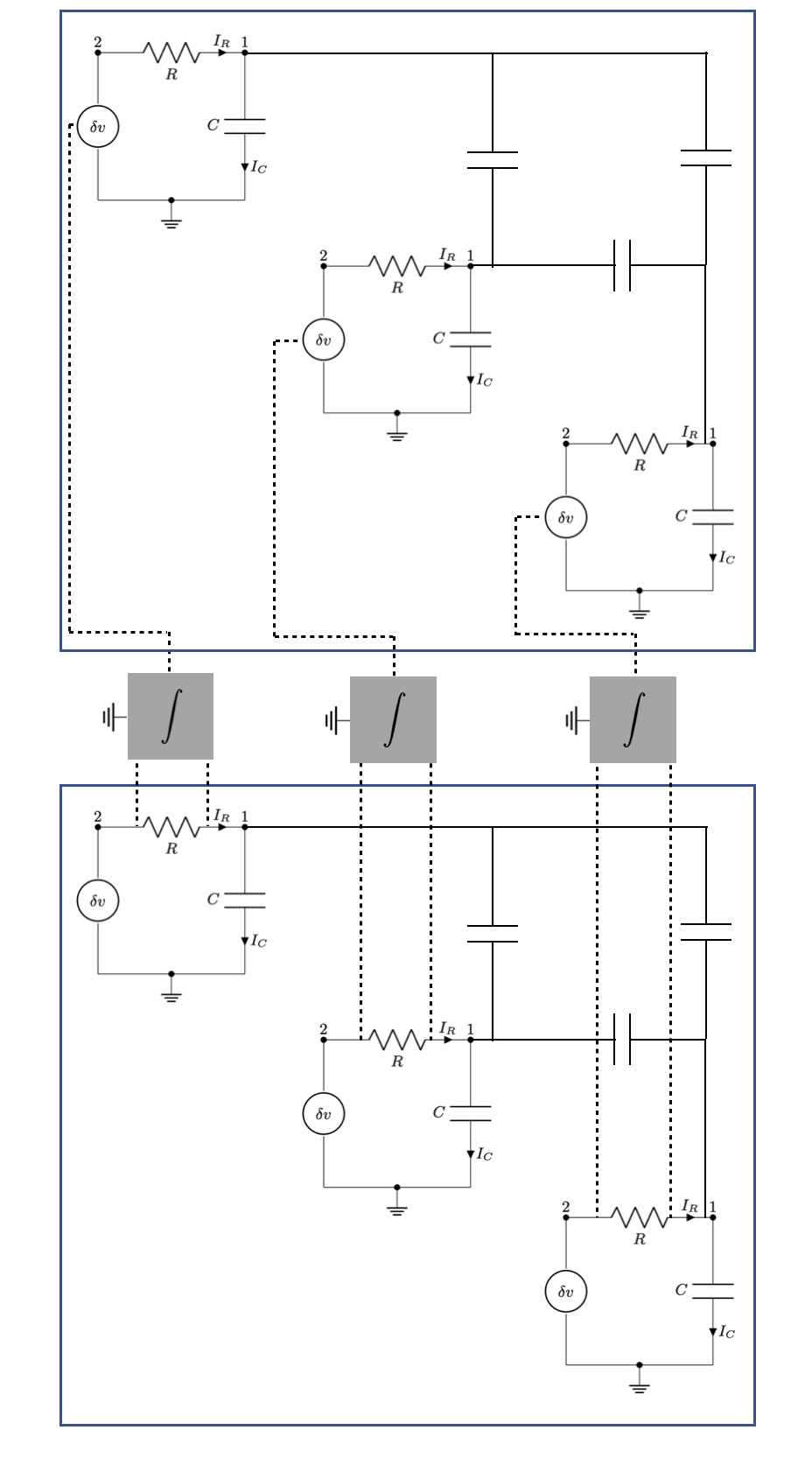}
    \caption{\textbf{Potential circuit diagram for matrix exponentiation hardware.} The lower subunit produces Gaussian noise, at the output of the integrators, with zero mean and a covariance matrix proportional to $A$. The top subunit directly simulates the desired stochastic differential equation in Eq.~\eqref{eq:specific_OU}.}
\label{fig:Circuit_Integrator}
\end{figure}

The thermodynamic matrix exponential algorithm requires a Gaussian noise source with a specified covariance matrix $\sigma$ (in this case $\sigma =\beta^{-1}( A + A^\intercal)$). While the noise could be generated by purely digital means, this may destroy the asymptotic speedup of the algorithm, as a Cholesky decomposition would be involved, which takes $O(d^3)$ time. We briefly suggest an alternative hardware-based method of generating the Gaussian noise which doesn't require any Cholesky decomposition.

Equation \eqref{langevin-equation} is called an overdamped Langevin equation, and can be implemented using an electrical circuit with $d$ nodes, each of which is coupled to all of the other nodes by a capacitor. Each node has a resistor, a capacitor, and a noise source (see Fig.~\ref{fig:Circuit_Main} in the main text or Fig.~\ref{fig:Circuit_Integrator} below). Associated with this circuit is a Maxwell capacitance matrix $\mathcal{C}$, whose elements are determined in the following way: let $C_i$ by the capacitance of the capacitor in node $i$ and $C_{ij}$ be the capacitance of the capacitor connecting nodes $i$ and $j$. Then the diagonal elements of $\mathcal{C}$ are given by $\mathcal{C}_{ii} = C_i + \sum_{j \neq i} C_{ij}$, while the off-diagonal elements are given by $\mathcal{C}_{i\neq j} = -C_{ij}$. The potential energy stored in all of the capacitors can be expressed as
\begin{equation}
    E =\frac{1}{2} V^\intercal \mathcal{C} V,
\end{equation}
where $V$ is the vector whose $i$th component is the voltage across the capacitor in the $i$th node. Note that at equilibrium $V$ is a random variable distributed according to Boltzmann distribution
\begin{equation}
    V\sim \mathcal{N}[0, \beta^{-1} \mathcal{C}^{-1}].
\end{equation}
This fact can be used to generate Gaussian samples \cite{coles2023thermodynamic}, however the covariance will proportional be the inverse of $\mathcal{C}$ rather than $\mathcal{C}$ itself, seemingly implying that a matrix inversion is necessary in order to sample from the correct distribution. Instead, we define a vector $\mathcal{Q} = \mathcal{C} V$, which is distributed as
\begin{equation}
    \mathcal{Q} \sim \mathcal{N}[0, \beta^{-1} \mathcal{C}].
\end{equation}
It is straightforward to see that the $i$th element of $\mathcal{Q}$ is the total amount of charge that has flowed through the resistor in the $i$th cell. Proceeding from the definition, we have
\begin{align}
    \mathcal{Q}_i &= C_i V_i + \sum_{j\neq i} C_{ij}( V_i -V_j) \\
   & =  C_i V_i + \sum_{j\neq i} C_{ij}V_{ij},
\end{align}
where we have defined $V_{ij} = V_i - V_j$, which is the voltage across the capacitor $C_{ij}$. Taking the time derivative gives
\begin{align}
    \frac{d\mathcal{Q}_i}{dt} &= C_i \frac{d V_i}{dt} + \sum_{j\neq i} C_{ij}\frac{d V_{ij}}{dt}\\
   & =  I_i + \sum_{j \neq i} I_{ij}\\
   & = I_{R_i}
\end{align}
where in the second line we defined $I_i$ as the current flowing through the capacitor $C_i$, and $I_{ij}$ to be the current flowing through capacitor $C_{ij}$, and in the third line (which follows from Kirchoff's current law) we defined $I_{R_i}$ to be the current through the resistor in the $i$th cell. Therefore samples of $\mathcal{Q}$ can be found by integrating the current through each resistor:
\begin{equation}
    \mathcal{Q}_i(t) = \int_0^t dt' \, I_{R_i}(t).
\end{equation}
The above integration can be performed by an analog integrator circuit at each node, which is fed the voltage across the resistor as an input. The integrated signal is then sent to a second thermodynamic device to be used as the noise source for the thermodynamic matrix exponential algorithm. This approach is illustrated in the circuit diagram in Figure~\ref{fig:Circuit_Integrator}.

\section{Comparison with Analog Algorithms}

\subsection{Contrasting Analog Algorithms with Thermodynamic Algorithms}

To get a better understanding of the advantages of thermodynamic computing, we wish to compare this paradigm to standard analog computing. This can help us understand the mechanism by which thermodynamic computing provides an advantage, including the notion of ``thermodynamic parallelism''.

Thermodynamic algorithms are characterized by the identification of computational targets with properties of the equilibrium probability distribution of a physical system. The evaluation of these computational targets is then carried out by allowing the system to come to equilibrium and then estimating properties of the equilibrium distribution (for example, its moments).

In contrast, traditional methods of analog computing have relied on measuring properties of a physical system at the \emph{trajectory} level rather than the \emph{distribution} level. That is, the solution to a problem is encoded directly in some (perhaps time-dependent) quantities that can be physically measured, rather than in the statistical properties of such quantities. Thus, a reasonable mathematical model for an (ideal) analog computer is an ordinary differential equation (ODE) solver, where some initial state $x(0)$ is deterministically evolved forward in time to some final state $x(t)$.

However, all analog computations are subject to noise of some kind. For example, thermal noise and shot noise are ubiquitous in electrical circuits, and hence they cannot be avoided completely. Thus, a mathematical model for a realistic analog computer necessarily accounts for sources of stochastic noise. This means that to compute some function to arbitrary accuracy with an analog computer will require some form of averaging, even if the algorithm is not based on the properties of the system's equilibrium distribution.

In other words, earlier analog algorithms have generally been limited to the use of first moments, whereas thermodynamic algorithms introduce the use of higher moments in computing various functions. With this in mind, in the next section we first describe a naive analog approach to computing the inverse of a matrix, and then a similar approach to computing the exponential of a matrix. These algorithms are limited to the use of first moments, and by the above reasoning we therefore do not consider them  to be thermodynamic algorithms.

\subsection{Naive Analog Algorithms for Matrix Inversion and Matrix Exponentiation}

\subsubsection{Matrix Inversion}

Suppose we have an analog device whose dynamics are described by the differential equation
\begin{equation}
\label{analog-inverse-ode}
    dx = - (A x - b) \, dt 
\end{equation}
for a symmetric positive definite matrix $A\in \mathbb{R}^{d\times d}$ and a vector $b\in \mathbb{R}^d$. As mentioned earlier, there will always be some stochasticity present in the system's dynamics, but in this case we do not describe the stochasticity directly via a stochastic differential equation, but instead will use a simplified model to capture heuristically the effects of noise on the solution. The solution $x(t)$ becomes arbitrarily close to $A^{-1} b$ for large times, so some sufficiently long time $T$ may be chosen such that the error at time $T$ is dominated by thermal noise rather than transient behavior. To describe the effects of thermal noise, we assume that at time $T$ the vector $x$ is distributed according to
\begin{equation}
    x(T) \sim \mathcal{N}[A^{-1} b, \sigma^2 \mathbb{I}],
\end{equation}
for some $\sigma>0$. Now note that the $i$th column of $A^{-1}$ is equal to $A^{-1} \hat{e}_i$, where $\hat{e}_i$ is the $i$th standard basis vector. Therefore the matrix $A^{-1}$ can be obtained by  the following protocol: first solve Eq. \eqref{analog-inverse-ode}, where $b = \hat{e}_1$, to obtain an estimate of $A^{-1}\hat{e}_1$. Repeat this process $N$ times and average the results to get a more accurate estimate of $A^{-1}\hat{e}_1$ (the necessary value of $N$ will be discussed later). Next repeat the previous steps for each of the other standard basis vectors, $\hat{e}_2, \hat{e}_3, \dots \hat{e}_d$. This process results in an estimate $M \approx A^{-1}$. We now consider the root-mean-square (RMS) Frobenius error of this solution, defined as
\begin{equation}
    \mathcal{E} = \sqrt{\left\langle \left \|A^{-1} - M \right \|_F^2\right \rangle}.
\end{equation}
Using the error model outlined above, we find
\begin{align}
    \mathcal{E} &= \sqrt{ \sum_{ij} \left\langle (A^{-1}_{ij} - M_{ij})^2 \right \rangle}\\
    & = \sqrt{ \sum_{ij} \text{Var}(M_{ij})} \\
    & = \frac{d \sigma}{\sqrt{N}}.
\end{align}
By the above analysis, in order to achieve some target error $\hat{\mathcal{E}}$, independent of dimension $d$, we must have $\sqrt{N}$ proportional to $d$, meaning $N \propto d^2$ (or larger). Finally, because the differential equation must be solved $O(d^2)$ times for each basis vector (of which there are $d$), the protocol requires an amount of time scaling as $O(d^3)$. However, the time requirements could be reduced in exchange for more physical resources by using a set of $d$ copies of the hardware to solve the separate differential equations simultaneously, for example.

\subsubsection{Matrix Exponentiation}
A similar strategy could be used to find the exponential of a matrix $A$. In this case, the device's dynamics are described by the differential equation
\begin{equation}
\label{analog-exponential-ODE}
    dx = - A x \, dt,
\end{equation}
and for some initial condition $x(0)$, the solution to the initial value problem (IVP) is given by
\begin{equation}
    x(t) = e^{-A t}x(0).
\end{equation}
Once again, we choose as an initial condition the standard basis vector $\hat{e}_1$, and after waiting an amount of time $\tau$, the state of the device may be measured to yield the vector $e^{- A \tau} \hat{e}_1$. Note that $e^{-A \tau}\hat{e}_i$ is the $i$th column of $e^{-A \tau}$, so if this process is repeated once for each standard basis vector, the matrix $e^{-A \tau}$ can be constructed from the solutions. Assuming the error is normally distributed about the correct solution, the same considerations apply as for the matrix inverse above. That is, we assume that the result read from the device is a random variable distributed as
\begin{equation}
    x(\tau) \sim \mathcal{N}\left[e^{-A \tau}x(0),\sigma^2 \mathbb{I}\right],
\end{equation}
and by the same logic used for the matrix inversion algorithm above, the RMS error will be
\begin{align}
    \mathcal{E} & = \frac{d \sigma}{\sqrt{N}},
\end{align}
where $N$ is the number of times the protocol is repeated before averaging. We see that this method of exponentiating a matrix requires time scaling with $O(d^3)$.

\section{Thermodynamic Parallelism}
We will find that the naive analog algorithms described above can be improved upon using a thermodynamic approach, and we ascribe this improvement to thermodynamic parallelism. In this section we explain the effects of thermodynamic parallelism conceptually using the example of matrix exponentiation.

First, imagine we implement the analog algorithm described above using $d$ separate analog devices, each of which solves Eq. \eqref{analog-exponential-ODE}. We describe the state of the collective system using a matrix $X$, whose $i$th column is the state vector of the $i$th device. At time $t=0$, the $i$th device is initialized to the standard basis vector $\hat{e}_i$, so we can write the collective initial condition as
\begin{equation}
    X(0)=\mathbb{I}.
\end{equation}
As time progresses, the state vector of each device evolves according to Eq. \eqref{analog-exponential-ODE}, so the dynamics of all of the devices together can be written as
\begin{equation}
\label{analog-matrix-ODE}
d X = -A X \, dt.
\end{equation}
This implies that at time $\tau$, the collective state of the devices is given by
\begin{equation}
X(\tau) = e^{- A \tau}X(0).
\end{equation}
Because the initial condition was chosen to be $X(0)=\mathbb{I}$, we have, as desired, $X(\tau)=e^{-A \tau}$. The key insight of thermodynamic parallelism is that the matrix differential equation \eqref{analog-matrix-ODE} can in fact be solved on a single device, if we consider the dynamics of the probability distribution of the device's state, rather than the state itself. In particular, the correlations between different components of the state vector will encode the elements of $X$.

Suppose we have a single device whose dynamics correspond to the following stochastic differential equation
\begin{equation}
    dx =- A x + \mathcal{N}[0,B dt],
\end{equation}
where $B\in \mathbb{R}^{d \times d}$ is symmetric and positive-definite. We consider the correlation matrix of the system, defined as
\begin{equation}
   C(t,s) =  \braket{[x(t) - \braket{x(t)}][x(s) - \braket{x(s)}]^\intercal}.
\end{equation}
To simplify the analysis, suppose that $\braket{x(0)}=0$, which implies that $\braket{x(t)}=0$ at all times. The correlation matrix becomes
\begin{equation}
   C(t,s) =  \braket{x(t)x^\intercal(s)}.
\end{equation}
We hold $s$ fixed, and write the change in $C$ corresponding to an infinitesimal change in $t$ as
\begin{align}
   d C(t,s) &=  \braket{(-A x(t) + \mathcal{N}[0,B dt])x^\intercal(s)}\\
   & = -A \braket{x(t) x^\intercal(s)} + \braket{\mathcal{N}[0,B dt]x^\intercal(s)}.
\end{align}
We assume that $t>s$, and in this case $x(s)$ cannot be correlated with the noise that is present at time $t$. Note that the the second term above is the mean of the inner product of two mean-zero uncorrelated random vectors, and therefore is zero. We arrive at
\begin{equation}
\label{thermo-correlation-ODE}
    dC(t,s) = -A C(t,s) dt.
\end{equation}
Note that Eqs. \eqref{analog-matrix-ODE} and \eqref{thermo-correlation-ODE} are formally identical when $C(t,s)$ is regarded as a function of $t$ and identified with $X(t)$. It is clear that for $\tau>0$ we have $C(t+\tau,t) = e^{-A \tau} C(t,t)$, so in order to recover the matrix exponential $e^{-A \tau}$, we must again have an initial condition $C(t,t) = \mathbb{I}$ for some $t$. By definition, $C(t,t)$ is simply the covariance matrix of the system $\Sigma$ at time $t$. After the system has come to equilibrium (which occurs after some suitably long time $T$ has elapsed), the covariance matrix is determined by the equation \cite{gardiner1985handbook}
\begin{equation}
    A \Sigma + \Sigma A^\intercal = B.
\end{equation}
We assume here that $A$ has positive real eigenvalues (which ensures that the above equation has a unique solution), and for simplicity we also assume that $A$ is symmetric, although this will be generalized in the following section. Therefore if we set $B= 2A$, we find that in the equilibrium distribution $\Sigma = \mathbb{I}$, or equivalently $C(t,t) = \mathbb{I}$ for large $t$, as desired. The above arguments show that the thermodynamic algorithm can be seen as a method of solving $d$ independent differential equations on a single device by looking at the dynamics at the level of the probability distribution, as opposed to the state of the system.

One detail that is not entirely clear from this analysis is how much time it takes to estimate the correlation matrix $C(t,s)$. If an amount of time scaling with $d^3$ were required to estimate the correlation matrix to arbitrary accuracy, then there would be no advantage of this method over the non-thermodynamic approach described earlier. As it turns out, though, the sampling complexity of estimating the correlation matrix scales at most with $d^2$, or in other words it is similar to the sampling complexity of the non-thermodynamic algorithm. This analysis is carried out in detail the section on convergence of the matrix exponential algorithm below. The advantage of the thermodynamic method can therefore be attributed to the fact that each ``sample" of the matrix $X$ is obtained constant time, instead of time scaling with dimension.

\subsubsection{The role of noise}
We should also clarify the role of noise in the thermodynamic algorithm. Because $A$ has eigenvalues with positive real part, the eigenvalues of $e^{-A t}$ have modulus less than one at all times $t>0$. This implies that, in the absence of noise, vectors which are different at time zero will become more similar as a result of the evolution (ie, the dynamics are contractive). This is easy to see as, given two vectors $x$ and $y$, for the noiseless dynamics we have
\begin{equation}
    \left|x(t) - y(t) \right| \leq \left|e^{-A t} \right|\cdot \left| x(0) - y(0)\right| \leq\left| x(0) - y(0)\right|.
\end{equation}
Another characterization of contractivity is that if the state of the system is a random variable, its Shannon entropy $S = -\int dx p(x) \ln p(x) $ decreases over time.

\end{appendix}
\end{document}